# Tetragonal mixed system $Cs_2CuCl_{4-x}Br_x$ complemented by the tetragonal phase realisation of $Cs_2CuCl_4$

*Natalija van Well\* [1,2], Claudio Eisele[1], Sitaram Ramakrishnan[1], Tian Shang[3], Marisa Medarde[3], Antonio Cervellino[4], Markos Skoulatos[5,6], Robert Georgii[5,6], Sander van Smaalen[1]*

[1] *Laboratory of Crystallography, University of Bayreuth, D-95447 Bayreuth, Germany*

[2] *Laboratory for Neutron Scattering and Imaging, Paul Scherrer Institute, CH-5232 Villigen, Switzerland*

[3] *Laboratory for Multiscale Materials Experiments, Paul Scherrer Institute, CH-5232 Villigen, Switzerland*

[4] *Swiss Light Source, Paul Scherrer Institute, CH-5232 Villigen, Switzerland*

[5] *Heinz Maier-Leibnitz Zentrum, Technische Universität München, Lichtenbergstr. 1, D-85747 Garching, Germany*

[6] *Physik Department E21, Technische Universität München, James-Franck-Str. 1, D-85747 Garching, Germany*

## ABSTRACT

Realisation of the tetragonal phase of $Cs_2CuCl_4$ is possible using specific crystal growth conditions at a temperature below 281K. This work deals with the comparison of the magnetic susceptibility and the magnetization of this new tetragonal compound with the magnetic behaviour of tetragonal $Cs_2CuCl_{2.9}Br_{1.1}$, $Cs_2CuCl_{2.5}Br_{1.5}$, $Cs_2CuCl_{2.2}Br_{1.8}$ and presents consistent results for such quasi 2-D antiferromagnets. Structural investigation at low temperature for $Cs_2CuCl_{2.2}Br_{1.8}$ shows no phase transition. The structure remains in the tetragonal symmetry $I4/mmm$. Furthermore, several magnetic reflections corresponding to the propagation vector $k = (0, 0, 0)$ are observed for this tetragonal compound through neutron diffraction experiments below the magnetic phase transition at $T_N = 11.3K$ confirming its antiferromagnetic nature.

## I. INTRODUCTION

Low-dimensional quantum spin systems reveal a wealth of fascinating phenomena in condensed matter[1]. Such systems include the $Cs_2CuCl_{4-x}Br_x$ mixed system, which exists in orthorhombic and tetragonal polymorphs[2]. In order to understand the influence of different $Cu^{2+}$ environments on the magnetic properties (with and without magnetic field) of the orthorhombic and tetragonal compositions of this mixed system, which is determined by the Cl/Br ratios, an investigation of $Cs_2CuCl_{4-x}Br_x$ is of interest.

The spin-½ antiferromagnetic Heisenberg model on a square lattice or anisotropic triangular lattice of the parent compounds $Cs_2CuCl_4$ and $Cs_2CuBr_4$ is used, to describe the dynamics of the spin degrees of freedom[3,4]. Regarding $Cs_2CuCl_{4-x}Br_x$, this model describes these well-known orthorhombic parent compounds as well as orthorhombic compositions between these two ones, which can be grown in the same orthorhombic symmetry *Pnma* in the full Br concentration range ($0 \leq x \leq 4$)[2,5]. The magnetic properties of this orthorhombic mixed system show two regions with long-range magnetic order (LRO) and two regions without magnetic order in the middle of the concentration range[5]. Long-range antiferromagnetic order occurs for $0 \leq x < 1.5$ below $T_N = 0.62K$ ($x = 0$) and for $3.2 < x \leq 4$ below $T_N = 1.4K$ ($x = 4$)[6,7]. In the orthorhombic mixed system, the changes in the tetrahedral $Cu^{2+}$ environment are characterized through different Cl and Br preferred occupations. This has a significant impact on the variations of the magnetic behaviour and its control[5].

Compounds of the tetragonal $Cs_2CuCl_{4-x}Br_x$ mixed system crystallize in the *I4/mmm* space group, and investigations on the polymorph compounds of the tetragonal phase of $Cs_2CuCl_{4-x}Br_x$ have been executed for the concentration range $1 \leq x \leq 2$[2,8]. In those tetragonal phases, the $Cu^{2+}$ environment consists of [CuX6] octahedra (X = Cl and/or Br). The octahedral coordination has strong Jahn-Teller distortions, which are elongated along the *a*- and *b*-directions. The tetragonal phase has a limited thermal stability and is irreversibly transformed into the orthorhombic phase, for example for tetragonal $Cs_2CuCl_{2.4}Br_{1.6}$ at $418K$[2]. The tetragonal-orthorhombic transition can be also activated by milling at room temperature[2]. A reverse orthorhombic-tetragonal transition cannot be achieved[2]. The magnetic behaviour of the tetragonal compounds can be described as quasi-2D antiferromagnets with an antiferromagnetic (AFM) transition at $T_N \approx 10K$ and a critical magnetic field of 1.5T at 2K, which is much smaller than expected for such compounds[8]. Prominent examples for magnetic behaviour of similar layer-type compounds are the tetragonal model systems $K_2CuF_4$ and $Rb_2CuCl_4$[9,10].

In this paper, the experimental details are presented in Sec. II. In Sec. III a description and the results of growing the new tetragonal polymorph of $Cs_2CuCl_4$, the structural characterization of selected tetragonal compounds of the mixed system ($Cs_2CuCl_{2.9}Br_{1.1}$, $Cs_2CuCl_{2.5}Br_{1.5}$ and $Cs_2CuCl_{2.2}Br_{1.8}$) and the investigation of the magnetic properties of the new tetragonal compound and the three other selected ones are shown, followed by conclusion and outlook.

## II.     EXPERIMENTAL DETAILS

**Crystal growth of tetragonal compounds of $Cs_2CuCl_{4-x}Br_x$ with Br concentration**

The tetragonal single crystals $Cs_2CuCl_{2.9}Br_{1.1}$, $Cs_2CuCl_{2.5}Br_{1.5}$ and $Cs_2CuCl_{2.2}Br_{1.8}$ were grown from aqueous solution using the evaporation method at room temperature[11]. For compounds with $1 \leq x \leq 2$, the crystalline reagents CsCl ($\geq$ 99.999%, Roth), $CuCl_2 \cdot H_2O$ (Analar Normapur, Merck) and $CuBr_2$ ($\geq$ 98%, Alfa Aesar) were used and the chemical reaction equations for the formation of the tetragonal phase have been applied[2]. Furthermore, tetragonal compounds with Br concentration were grown at different temperatures between 277K and 297K[11]. The colour of crystals with $1 \leq x \leq 2$ is generally dark red. The chemical compositions of the samples were determined through EDX analysis, using a Zeiss Leo 1530. In the calculation of the chemical composition of the compounds with different Br concentrations, the ratio value of Cl/Br was utilized. To facilitate the presentation in the following, a nominal value for this Br concentration is used in this paper.

**X-ray diffraction**

X-ray powder diffraction (PXRD) was performed at the Powder Diffraction station of the Materials Sciences Beamline (X04SA-MS) at the Swiss Light Source (SLS) at the Paul Scherrer Institute (PSI) in Villigen[12]. The powder sample was investigated together with diamond-standard, enclosed in a capillary with a diameter of 0.3 mm, which was placed in a Janis flow-type cryostat at different temperatures between 295K and 4K. A wavelength of $\lambda = 0.775369$Å was selected, and a Microstrip Mythen-II detector was used, which allowed for high counting rates while maintaining the high spatial resolution.

The Rietveld method was utilized, to extract the structure information from the PXRD pattern. To fit the observed PXRD data of the investigated tetragonal compounds, the Rietveld method with three phases was applied: (1) tetragonal structure model oriented parallel to the beam, (2) tetragonal structure model oriented perpendicular to the beam, and (3) cubic structure model for the diamond standard. Two tetragonal phases were used to correct a sample displacement of the peak-shape profile of the powder diffraction data[13]. Despite the required capillaries in the Debye-Scherrer geometry, a preferred orientation was detected, which originates from the plate-like form of the powder crystallites and was described by the use of symmetry-adapted spherical harmonics. The Thompson-Cox-Hastings pseudo-Voigt function was important to define the peak-shape profile. The background was modelled by cubic splines interpolation of manually selected data points. The background points are the same for the three phases and were kept fixed. PXRD data were refined using the FULLPROF program[14].

**Magnetic susceptibility measurements**

The magnetic susceptibility was measured using a Quantum Design MPMS magnetometer in a temperature range from 1.8K to 300K. The samples were oriented with the *c*-axis perpendicular and parallel to the magnetic field. The magnetic moment from the sample holder together with GE varnish, used to fix the sample, was measured at the same temperature range. The background contribution of the sample holder with varnish was taken into account with respect to the data shown in Figure 5. All experimental magnetic data were corrected for diamagnetism of the constituent atoms (Pascal's tables)[15].

**Neutron diffraction**

Single crystal neutron diffraction experiments were carried out on the cold neutron powder diffractometer DMC at SINQ of Paul Scherrer Institute (PSI) in Villigen, Switzerland in single crystal mode[16,17]. A monochromatic neutron beam with wavelength λ = 2.458Å was produced through a $PG_{002}$-monochromator. A PG-filter was used to remove higher order contamination. Normal beam geometry was used. The diffracted signal has been collected with an area detector. The crystals were cooled down to 1.5K. Single crystal neutron diffraction measurements were performed also on the cold-three-axes spectrometer MIRA at FRM II of Heinz Maier-Leibnitz Zentrum (MLZ) in Garching, Germany[18]. For the elastic measurements, a wavelength of 4.488Å (ki = 1.4A$^{-1}$) was used by means of a $PG_{002}$- monochromator. A Be-filter and a single $^3$He-tube detector were used. The crystals were cooled down to 3.5K.

## III. RESULTS AND DISCUSSIONS

**Crystal growth of new tetragonal polymorph of $Cs_2CuCl_4$**

The tetragonal polymorph of $Cs_2CuCl_4$ can be grown below 281K. The crystallization is performed from the quasi-ternary system $CsCl$-$CuCl_2$-$H_2O$. The phase diagrams at 291K, 298K and 323K were extensively exploited and published in the past[19]. These investigations show that the anhydrous phases in this system crystallize only above 293K, assuming a constant temperature profile. For the crystal growth of tetragonal $Cs_2CuCl_4$, a periodically changing temperature profile with an amplitude of 0.3K and a period of 8 hours was used.

As a starting point for growing the tetragonal crystals, molar ratios from 2:1 to 6:1 of CsCl and $CuCl_2$ were selected. In the beginning of the experiment, a saturated solution was prepared, which then evaporated at a temperature of 281K with different rates. The evaporation rate is important for the quality and the growth time of the crystals. The time for the crystallization of this mixed system is between four weeks (rather small crystals) and seven months (rather large crystals and crystals of higher quality), depending on the evaporation rate. The volume of the crystallized material is small. The obtained crystals are yellow (transparent) with a quadratic surface.

It should be noted that an irreversible phase transition occurs for tetragonal $Cs_2CuCl_4$ at 289K into the orthorhombic phase and the transparency of the crystals disappears. In addition, whereas the crystal growth with the molar ratios 2:1 and 3:1 of CsCl and $CuCl_2$ generally result in the formation of the two phases $Cs_2CuCl_4$ and $Cs_3Cu_3Cl_8OH$, only tetragonal $Cs_2CuCl_4$ is grown using molar ratios from 4:1 to 6:1.

The chemical composition of the samples was determined by way of energy dispersive X-ray analysis (EDX) and showed the following results in at.%: Cs-29.51±0.21, Cu-15.03±0.21, Cl-55.46±0.15. This outcome is consistent with the chemical formula of $Cs_2CuCl_4$. In addition, a tetragonal phase has previously been reported for $Cs_2CuCl_4 \cdot 2H_2O$, whose crystals show a blue colour [19]. Single crystals of the mixed system are shown in Figure 1.

The conoscopic interference patterns of $Cs_2CuCl_4$ and $Cs_2CuCl_{2.2}Br_{1.8}$ show unique optical uniaxial ones (see Figure 1 b) and d)). This result indicates that the structure of these crystals is tetragonal[20]. However, the structure investigation of the single crystals of the tetragonal phase of $Cs_2CuCl_4$ with x-ray diffraction revealed difficulties preparing the crystals for the structure investigation, because the phase transition from the tetragonal into the orthorhombic phase occurs very quickly at 289K. In addition, milling of this compound at low temperatures for powder diffraction (around 80K) also changes the tetragonal phase of these crystals into the orthorhombic phase.

In general, the crystals of this tetragonal mixed system are grown as thin layers, which are shown in Figure 2a). The smallest measured thickness is around 13.06nm (see Figure 2b). Therefore, this compound can be called a nano-compound due to its structural layer configuration. Growing of the layers is such that they do not exactly order to each other. This is another difficulty for the single crystal structure investigation of such compounds. Such ordering results in a broadening of reflections and/or additional reflections, which cannot be traced back to the compound´s structure, but to the overlap of the thin layers of the composition. Consequently, such particularities raise difficulties for the structure solution through single crystal diffraction.

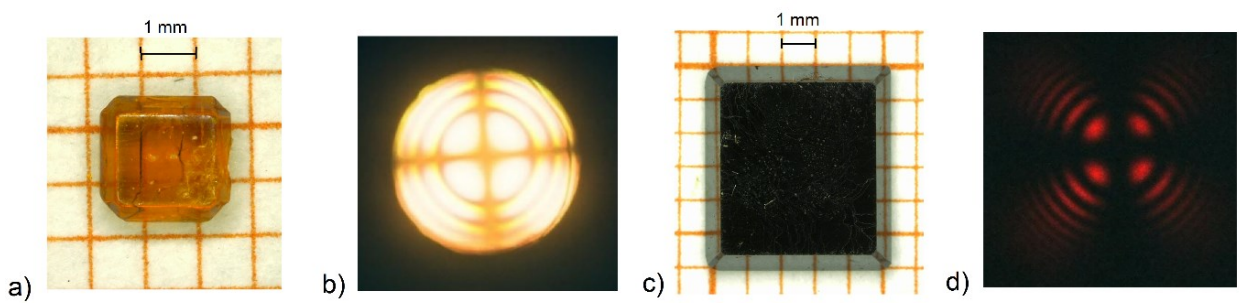

**Figure 1:** Single crystals of the tetragonal phase of a) $Cs_2CuCl_4$ and c) $Cs_2CuCl_{2.2}Br_{1.8}$; conoscopic interference pattern with the optical axis parallel to (*0 0 1*) for b) $Cs_2CuCl_4$ and d) $Cs_2CuCl_{2.2}Br_{1.8}$

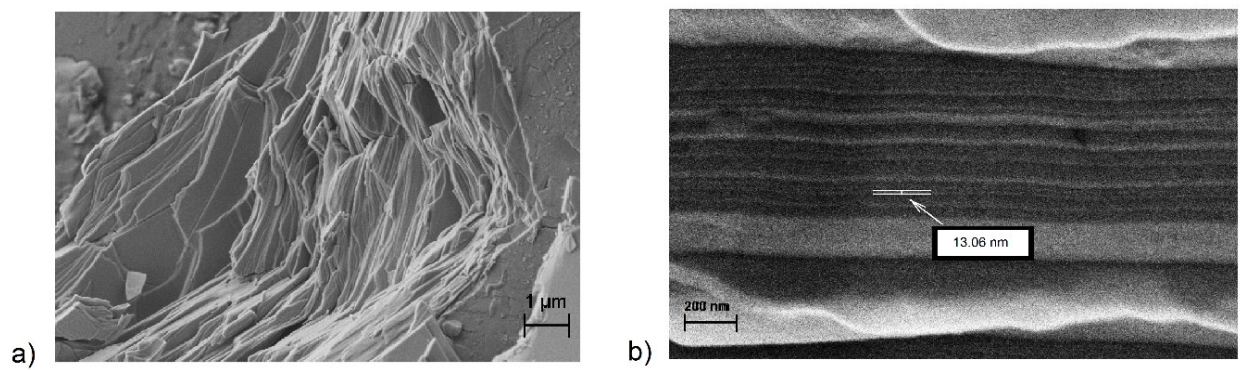

**Figure 2:** Electron microscopy pictures of the morphology of the investigated tetragonal crystals for a) an example of the fracture point of the $Cs_2CuCl_{2.2}Br_{1.8}$, b) side view showing the smallest layer thickness observed

For the tetragonal crystals with Br concentration ($Cs_2CuCl_{2.9}Br_{1.1}$, $Cs_2CuCl_{2.5}Br_{1.5}$ and $Cs_2CuCl_{2.2}Br_{1.8}$), the best alternative to investigate its structure is PXRD, as additional reflections from the overlap of the thin layers are not shown. The powders of the fore-mentioned compounds were produced at around 80K using a Retsch Cryomill. At room temperature, the milling process causes tetragonal to orthorhombic phase transition[2].

**Crystal structure determination at low temperature**

The low temperature crystal structure is important to describe and understand the magnetic properties of the investigated compounds of the mixed system. As already described above, an investigation of the structure of the new tetragonal $Cs_2CuCl_4$ with single crystal x-ray diffraction or PXRD is not possible. Therefore, in the following, only the results of investigations of Br-doped compositions of tetragonal $Cs_2CuCl_{4-x}Br_x$ are presented. At low temperatures, investigations were carried out on $Cs_2CuCl_{2.2}Br_{1.8}$ in order to draw conclusions on the relationships between structure and properties from the interplay of the results of all tetragonal compositions of this mixed system investigated.

Figure 3 shows a detailed view of the environment of $Cu^{2+}$ in $Cs_2CuCl_{2.9}Br_{1.1}$. The Cu atoms have an octahedral environment with Cl and/or Cl/Br atoms depending on the Br concentration, where all octahedrons are distorted. In such octahedrons, the distortion results from the environment with different Cl/Br atoms, which are overlapped with the Jahn-Teller-distortion. This results in an elongated distortion in the *a*- and *b*-directions for such $Cu^{2+}$ environments. In analogy, the compositions of the comparable $Rb_2CuCl_{4-x}Br_x$ also show this form of elongated distortion[21]. The analysis of the synchrotron data of $Cs_2CuCl_{2.9}Br_{1.1}$ confirms the preferred occupation of the Br-atoms on the halogen sites X2, as illustrated in Figure 3. This is already apparent from the development of the lattice parameters with x, shown in Table 1. Going from x = 1 to 2, the relative change occurs along the *c*-direction. This is similar to the relative change of *a=b*.

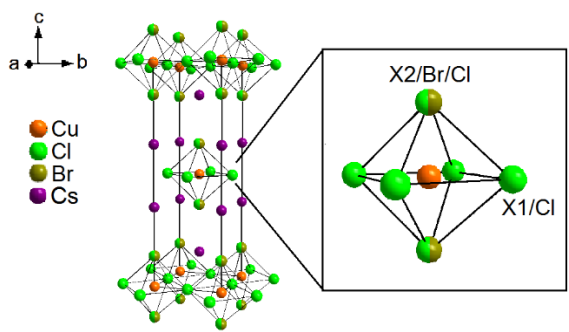

**Figure 3:** Perspective view of $Cs_2CuCl_{2.9}Br_{1.1}$ along the *a*-axis, in dark green: crystallographic positions partially occupied by Br

Looking on the octahedron and its location within the crystal structure, it is clear that the X2-site will have the strongest influence on the *c*-parameter of the tetragonal crystal structure as depicted in Figure 3. PXRD was measured and analysed for tetragonal $Cs_2CuCl_{2.9}Br_{1.1}$ and $Cs_2CuCl_{2.5}Br_{1.5}$ at 295K and for tetragonal $Cs_2CuCl_{2.2}Br_{1.8}$ from 295K to 4K. The structure of these compounds has been characterized in detail. For these compounds, a partially ordered distribution of the halogen components with respect to the preferred crystallographic positions was suggested[2]. The results of PXRD and EDX are summarized in Table 1. The Br concentrations show some differences, but are very close to the nominal values.

**Table 1:** Structure refinement results (*I4/mmm*) for three Br concentrations with PXRD and EDX; lattice parameter at 295K of the three investigated compositions and at 4K for $Cs_2CuCl_{2.2}Br_{1.8}$

|  | $Cs_2CuCl_{2.9}Br_{1.1}$ | $Cs_2CuCl_{2.5}Br_{1.5}$ | $Cs_2CuCl_{2.2}Br_{1.8}$ | |
|---|---|---|---|---|
| Nominal chemical Composition | x=1.1 | x=1.5 | x=1.8 | |
| EDX results * for Br concentration | x=1.13(2) | x=1.46(2) | x=1.76(2) | |
| PXRD results for Br concentration | x=1.15(5) | x=1.56(5) | x=1.68(5) | |
| Temperature [K] | 295 | 295 | 295 | 4 |
| a=b [Å] | 5.25335(2) | 5.26606(3) | 5.27828(3) | 5.22159(6) |
| c [Å] | 16.55410(7) | 16.6662(1) | 16.74032(9) | 16.6529(2) |
| Volume [Å$^3$] | 456.855(4) | 462.177(5) | 466.389(4) | 454.042(2) |
| Z; $\rho_{calc}$ [g cm$^{-3}$] | 2; 3.799 | 2; 3.883 | 2; 3.882 | 2; 3.996 |
| Occupation of X1 Cl1 - 0 0.5 0 Br1 – 0 0.5 0 | 0.9696(24) 0.0304(24) | 0.9264(32) 0.0736(32) | 0.9648(24) 0.0352(24) | 0.9656(32) 0.0344(32) |
| Occupation of X2 Cl2 – 0 0 z Br2 – 0 0 z | 0.4536(24) 0.5464(24) | 0.2944(32) 0.7056(32) | 0.2032(24) 0.7968(24) | 0.1888(40) 0.8112(40) |
| Distances [Å] Cu-Cl1/Br1 Cu-Cl2/Br2 | 2.626675(10) 2.4336(8) | 2.633030(15) 2.4696(7) | 2.639140(15) 2.4841(7) | 2.61080(3) 2.4708(10) |
| $R_p$; $R_{wp}$; $R_{Bragg}$ % | 1.705; 2.474; 4.134 | 1.896; 2.677; 6.804 | 1.618; 2.295; 5.232 | 1.628; 2.548; 6.962 |

(*) Average values of Br concentration from different measurements

CCDC 1892901, 1892903, 1894106 and 1894107 X-ray crystallographic files in CIF format.

The values of the partial occupancy of these positions after refinement are shown in Table 1. The crystallographic position X1 is occupied preferably by Cl, and only a small amount of Br. The X2 position is generally occupied by both, Cl and Br. On this position, a systematic change from Br to Cl takes place. The different distances of Cu-Cl and Cu-Br in the $Cu^{2+}$ environment give rise to a distortion in the octahedron.

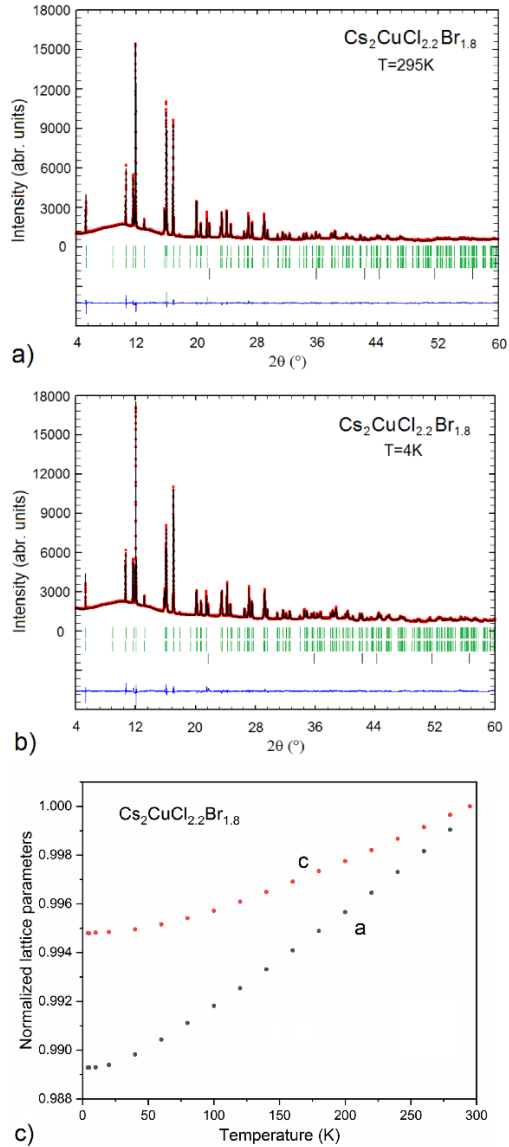

**Figure 4:** Powder diffraction of C$_2$CuCl$_{2.2}$Br$_{1.8}$ at: a) 295K and b) 4K, in green: the *h k l* of the space group *I*4/*mmm* and *Fd*-3*m* – in black, and in blue: the difference line; all measurements were executed with diamond - standard powder; c) normalized lattice parameter depending on the temperature

The PXRD data at different temperatures for the concentration x=1.8 are shown in Figure 4. The tetragonal structure *I*4/*mmm* of the crystals remains stable down to 4K without any indication of a structural phase transition between 295K and 4K. The obtained structural parameters as a function of temperature are shown in Figure 4c). The normalized lattice parameters for the *c*-axis are slightly smaller than for *a=b*, which indicates that there is an obvious change in *a*- and *b*-directions. The analysis of the octahedra distortion at different temperatures shows the tendency that at low temperatures (i.e. at 4K) the octahedron is slightly elongated along the *c*-direction compared to the octahedron distortion of Cs$_2$CuCl$_{2.2}$Br$_{1.8}$ at 295K.

**Magnetic properties**

This section deals with the comparison the magnetic properties of the new tetragonal composition to the Br-doped tetragonal compositions and shows correlations of the magnetic susceptibility and magnetization of the investigated compositions.

The magnetic susceptibility measurements show that the magnetic properties are dominated by quasi-2D interactions[8]. As shown in Figure 5a), the value of $T_N = (8.96 \pm 0.50)$K is determined from the susceptibility of $Cs_2CuCl_4$ and indicates AFM behavior. The compounds with increasing Br concentrations show the value of $T_N$ in the susceptibility at higher temperatures. A Br concentration with x = 1.8 results in $T_N = (11.26 \pm 0.50)$K. The magnetic susceptibility investigations also reveal a difference between the zero-field-cooled and field-cooled curves of the investigated compounds (see Figure A in supporting information). The Curie-Weiss temperature $\theta_{cw}$ was extracted from the temperature dependence of the inverse susceptibility $X^{-1}_{mol}(T)$ between 150K and 280K for tetragonal $Cs_2CuCl_4$ (see Figure 5b)). The fit resulted in C = 5.231(1)Km³·mol⁻¹ and $\theta_{cw} = (35.64 \pm 0.35)$K. The positive value of $\theta_{cw}$ implies that the ferromagnetic (FM) interaction is predominant.

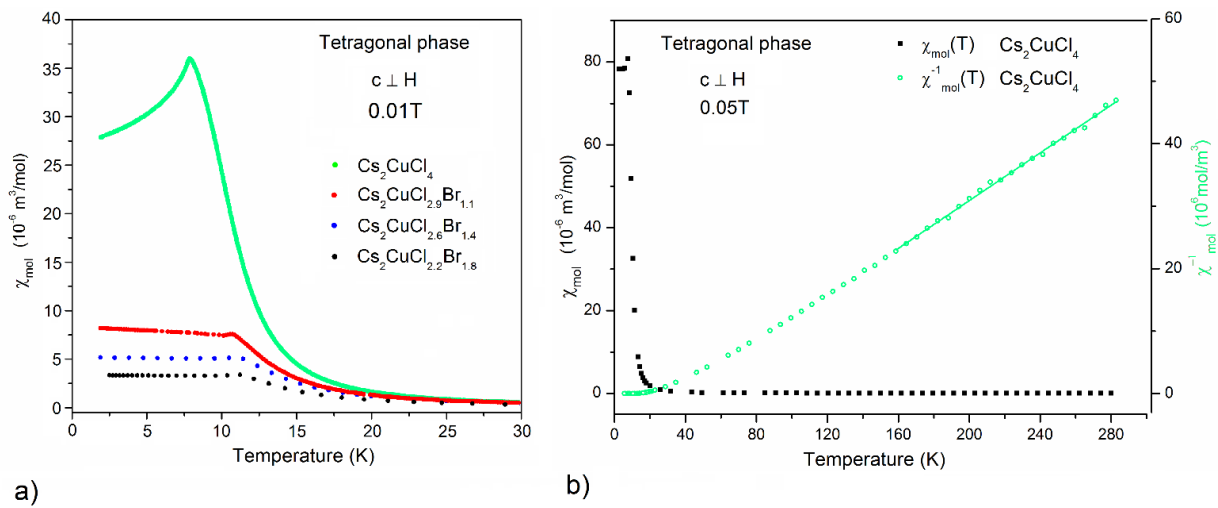

**Figure 5:** Magnetic susceptibility $X_{mol}(T)$ vs. T and inverse susceptibility $X^{-1}_{mol}(T)$ depending on the temperature in the magnetic field: a) detailed view of the susceptibility maximum for the three investigated tetragonal compounds at 0.01T and b) of tetragonal $Cs_2CuCl_4$ at 0.05T

A comparison with literature indicates that for the Curie-Weiss temperature a large difference of $Cs_2CuCl_4$, which was reported by H.-J. Kim et al. ($\theta_{cw} = (3.3 \pm 0.6)$K) and that of the investigated tetragonal compound ($\theta_{cw} = (35.64 \pm 0.35)$K) exists[22]. Nevertheless, this difference is attributed to the different applied temperature regions (used 150K – 280K vs 20K – 300K by H.-J. Kim et al.)[22]. In comparison to the tetragonal phase of $Cs_2CuCl_4$, the Curie-Weiss temperature of the orthorhombic phase of $Cs_2CuCl_4$ is in the region between 3.5K and 7.1K[23-25]. For tetragonal $Cs_2CuCl_{2.6}Br_{1.4}$, the fit resulted in C = 5.077(1)Km³·mol⁻¹ and $\theta_{cw} = (36.83 \pm 0.35)$K, which are comparable to the tetragonal phase of $Cs_2CuCl_4$.

Figure 5a) presents that i) the value of $T_N$, determined from the susceptibility, is shifted to a higher temperature with an increase of the Br concentration, and also that ii) the value of $X_{mol}(T_{max})$ is reduced as the Br concentration increases. This indicates that the magnetic coupling is modified most likely between the layers. The details still have to be investigated. Upon application of a magnetic field, $T_N$ shifts to lower temperatures. For $Cs_2CuCl_4$, this results in $T_N = (8.96 \pm 0.50)$K in a magnetic field 0.01T and in $T_N = (7.65 \pm 0.50)$K in a magnetic field 0.05T. In comparison to the x=1.5 composition, this value is $T_N = (10.47 \pm 0.50)$K in 0.01T and $T_N = (6.66 \pm 0.50)$K in 1T. For the x=1.8 composition, a similar shift was reported by Cong et al[8]. The value of $(X_{mol}(T)*T)^{1/2}$ increases depending on the temperature, with the maximum reached at $(8.64 \pm 0.50)$K for tetragonal $Cs_2CuCl_4$. Such increasing values show that a dominant FM coupling exists (see Figure B in the supporting information). After having reached the maximum, the value of $(X_{mol}(T)*T)^{1/2}$ decreases rapidly upon reducing the temperature. The effective magnetic moment of the $Cu^{2+}$ ions was determined at 1.96(4)$\mu_B$ perpendicular to the c-axis for the compound $Cs_2CuCl_4$, which is slightly larger than the spin-only value of 1.732$\mu_B$ for $Cu^{2+}$ ions. The value of the effective magnetic moment and the temperature independent behavior of $\mu_{eff}$ between 150K and 280K indicates an octahedral environment of $Cu^{2+}$.

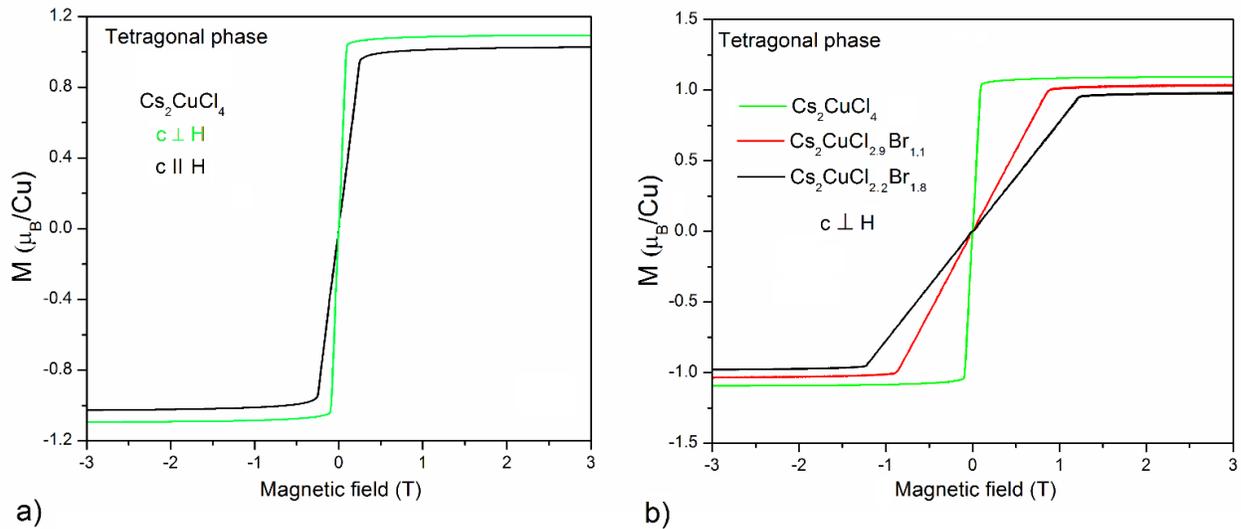

**Figure 6:** Field dependence of the magnetization of the tetragonal phase at $T = 2$K for a) $Cs_2CuCl_4$ with the *c*-axis parallel and perpendicular to the magnetic field and b) $Cs_2CuCl_4$, $Cs_2CuCl_{2.9}Br_{1.1}$ and $Cs_2CuCl_{2.2}Br_{1.8}$ with the *c*-axis perpendicular to the magnetic field

This effective magnetic moment confirms the orbital moment contribution of the $Cu^{2+}$ ions in such distorted octahedrons[26-28].

Figure 6a) shows the anisotropy of the magnetization along different directions for the investigated compound. The magnetization increases linearly with the magnetic field, which is applied perpendicular to the *c*-axis (maximum value $B_s = 0.1$T). The saturated magnetization is $\mu_s = 1.09\mu_B$. This results in the perpendicular Lande-factor $g_{senk} = 2.18$. In case that the c-axis is parallel to the magnetic field, the value of the saturated magnetic field is $B_s = 0.25$T and $\mu_s = 1.02\mu_B$. For the parallel Lande-factor $g_{par}$, we have calculated avalue of 2.05. It becomes obvious that the orientation of the magnetic moments in the *ab*-plane needs a smaller $B_s$ than the orientation of the magnetic moments along the *c*-axis. In comparison to the tetragonal phases $Cs_2CuCl_{2.2}Br_{1.8}$ and $Cs_2CuCl_{2.9}Br_{1.1}$, $B_s$ for the tetragonal phase $Cs_2CuCl_4$ is smaller. In Figure 6b), the results of the magnetization for these compositions at $T = 2$K are summarized. The magnetization increases linearly with the magnetic field up to saturation ($B_s$). It is slightly different for the investigated compositions: for x=1.1 $B_s$= 0.91T and for x=1.8 $B_s$ = 1.25T. Therefore, the Br concentration is clearly the reason for the shift of $B_s$. In general, for small fields, the spins align antiferromagnetically in the *ab*-plane. For strong fields, the spins develop a uniform component along the field and are thus canted out of plane[29]. In Figure 6b), different values of the $B_s$ for the corresponding Br concentrations can be seen, which are attributed to canting[30,31]. Hence, with different Br concentrations, the canting angles also change. The reason for that is the variation of the $Cu^{2+}$ environment, because the X2 crystallographic position in Cu-octahedra is partially occupied by Br depending on the concentration. Therefore, the exact determination of the magnetic structure will contribute to the determination of these interdependencies. To clarify the ground state of the tetragonal phase of the investigated compounds, we have measured reciprocal space maps for the composition $Cs_2CuCl_{2.5}Br_{1.5}$ (Figure 7). This gives an overview of the magnetic reflections at 1.5K.

Figure 7a) shows nuclear reflections in the (*h 0 l*) plane at 20K, and Figure 7b) displays the half integer magnetic reflections between the nuclear ones at 1.5K. The very small additional reflections near the nuclear reflections at both temperatures show that the layers in this compound are not exactly ordered to each other.

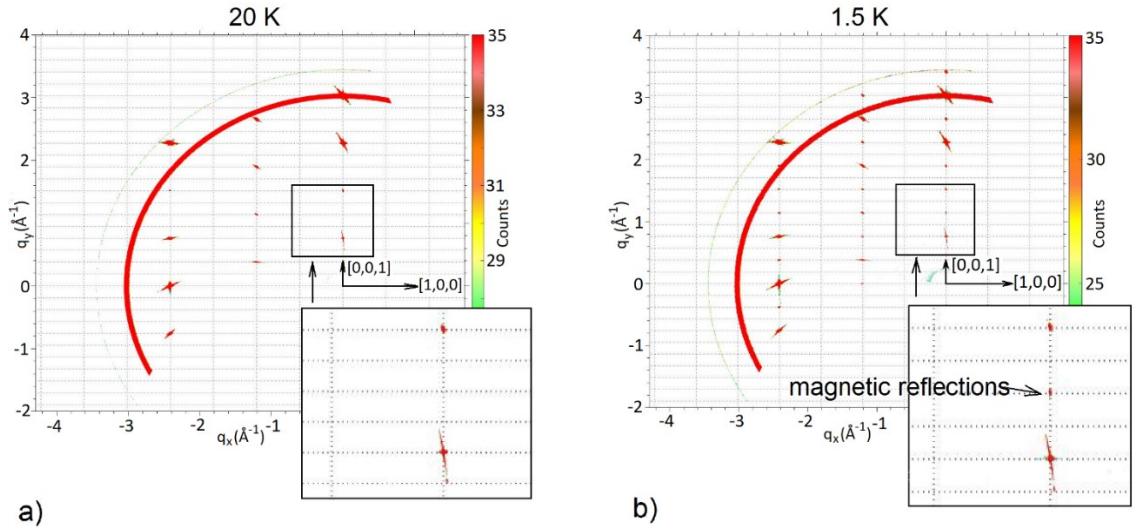

**Figure 7:** a) Zero-field intensity map of tetragonal $Cs_2CuCl_{2.5}Br_{1.5}$ in the (*h 0 l*) plane at a) 20K and b) 1.5K; the step ($\Delta\phi$) in these measurements was 0.2°

Figure 8a) presents the schematic reciprocal space map of the measured magnetic reflections in the (*h k 0*) and the (*h 0 l*) plane of tetragonal $Cs_2CuCl_{2.5}Br_{1.5}$. As a next step, we have determined the temperature dependence of the integrated intensities of one of them ((*1 0 0*), see Figure 8b)). Scans through the magnetic reflection with increasing temperature illustrate that the intensity of the magnetic reflection decreases by approaching the Neel temperature $T_N$, and disappears beyond it, confirming its magnetic nature.

The AFM ordering temperature of tetragonal $Cs_2CuCl_{2.2}Br_{1.8}$ is $T_N$ = 11.300(4)K, determined by a fit of power-law behaviour (see Figure 8c). This result is in very good agreement with the susceptibility measurement. Note also that the positions of the magnetic reflections don`t change with the temperature, indicating the absence of changes in the propagation vector, and that magnetic reflections occur in forbidden positions for the nuclear reflections of the space group *I*4/*mmm*. In particular, they violate the *I*-centering extinction condition $h + k + l = 2n$.

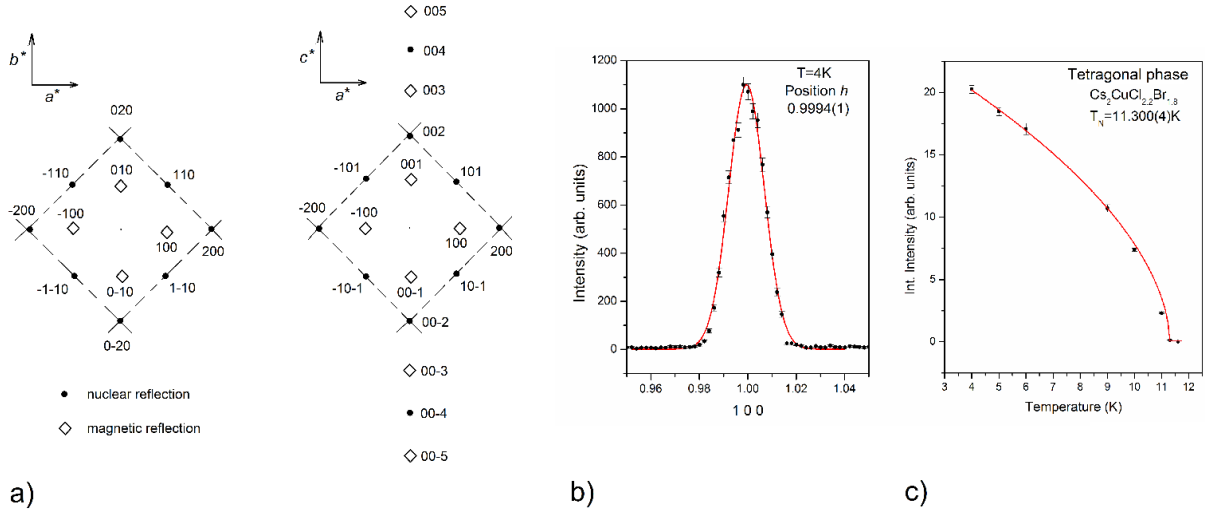

**Figure 8:** Tetragonal $Cs_2CuCl_{2.2}Br_{1.8}$: a) position of magnetic and nuclear reflections in *ab*- and *ac*-planes; b) intensity of the (*1 0 0*) magnetic peak; c) temperature dependence of the integrated intensity for the (*1 0 0*) peak and fit of power-law behavior marked in red

This means that the magnetic cell is not *I*-centered. 12 magnetic reflections for the compound $Cs_2CuCl_{2.2}Br_{1.8}$ in different planes have been measured, which are not sufficient for a complete determination of the magnetic structure of this compound. Nevertheless, the ground state, was determined as AFM. The orientation of the spins in the *ab*- and *ac*-planes could not yet be clarified.

IV. CONCLUSION AND OUTLOOK

Tetragonal $Cs_2CuCl_{4-x}Br_x$ with $1 \leq x \leq 2$ was complemented together with the Br-free realisation of tetragonal phase of $Cs_2CuCl_4$. The optical characterisation for this new tetragonal phase of $Cs_2CuCl_4$ shows a similar conoscopic interference pattern like that for other established tetragonal mixed compounds.

The structure investigations at different temperatures from 295K to 4K with PXRD for the tetragonal composition $Cs_2CuCl_{2.2}Br_{1.8}$ do not show any phase transition in the investigated temperature region. The structure still remains *I*4/*mmm* down to low temperatures.

The magnetic behaviour of the tetragonal phase $Cs_2CuCl_4$ fits very well to that of the tetragonal mixed system. The magnetization shows an anisotropy between perpendicular and parallel directions to the magnetic field.

The shift of $B_s$ of the Br-doped compounds depends on the doping concentration and increases with increasing Br concentration. The magnetic behaviour remains AFM down to 1.5K. One of the key points for understanding the magnetic structure of the investigated tetragonal compounds is, to clarify, whether there is a relationship between the spin ordering and the crystallographic ordering of the [CuX6] octahedra. To answer this question, single crystal neutron diffraction measurements are envisaged.

The Br concentration, which only slightly changes the distances between the planes, changes the $Cu^{2+}$ environment to a greater extent. The $Cu^{2+}$ environment consists of [CuX6] octahedrons, which are responsible for the composition of the tetragonal phase of $Cs_2CuCl_{4-x}Br_x$, which is composed of Cl- and Br-atoms. In addition to the Jahn-Teller distortion of the octahedron in this system, there is an additional distortion due to the different ligands that make up the octahedron. In this case, the magnetic behaviour reflects the change of the $Cu^{2+}$ environment. With increasing Br content, $B_s$ also increases, as measured perpendicular to the planes.

It seems that a reduction of the distances between the layers, for example by pressure, leads to FM order. However, it is known from literature that applied pressure results in a phase transition from a FM to an AFM interaction and to an increase of $T_N$ or $T_C$[10]. Therefore, this phenomenon has to be further investigated. Not only changes of the material properties, based on the variation of the distances between the planes, but also variations of the $Cu^{2+}$ environment give an insight into the correlations of FM and/or AFM interactions for this kind of compounds.

The understanding of these interactions leads to new parameters in the development of new materials of this material class. To date, it has not yet been determined, which structural characteristics between the planes lead to a FM interaction in $Cu^{2+}$.

**Conflicts of interest**

The authors declare no conflict of interest.

**Supporting Information**.

Figure A: Difference between the zero-field-cooled and field-cooled curves of the magnetic susceptibility measurements.

Figure B: The value of $(X_{mol}(T)*T)^{1/2}$ depending on the temperature

**Corresponding Author**

* Dr. Natalija van Well, University of Bayreuth, Universitaetsstr. 30, D-95447 Bayreuth,

    E-Mail: natalija.van-well@uni-bayreuth.de

**Author Contributions**

The manuscript was written through contributions of all authors.

**Funding Sources**

Deutsche Forschungsgemeinschaft through the research fellowship for the project WE-5803/1-1 and WE-5803/2-1, SNF under grant number 206021_139082 and University of Bayreuth through fellowship A 4576 – 1/3.

ACKNOWLEDGMENT

The authors thank W. Asmuss, C. Krellner, F. Ritter, P. Puphal, P. T. Cong, B. Wolf, M. Lang from Goethe-University, Frankfurt am Main, M. Heider from Bayerisches Polymerinstitut, Bayreuth, L. Keller and Ch. Rüegg from Laboratory for Neutron Scattering and Imaging (PSI), Villigen for fruitful discussions, and E. Canevet from Laboratory for Neutron Scattering and Imaging (PSI), Villigen for his support during the experiment at the DMC and the analysis of the data and D. Sheptyakov from Laboratory for Neutron Scattering and Imaging (PSI), Villigen for his support with the refinement of the synchrotron data. The authors also thank B. Pedersen for the orientation of the crystals on the neutron-laue diffractometer — RESI at FRM II of Heinz Maier-Leibnitz Zentrum (MLZ) in Garching. The synchrotron x-ray experiments were executed on X04SA-MS at SLS of PSI, Villigen, Switzerland. The neutron diffraction experiments were performed on DMC at SINQ of PSI, Villigen, Switzerland and MIRA at FRM II of MLZ, Garching. This work was supported by Paul Scherrer Institute, Heinz Maier-Leibnitz Zentrum and Physik Department E21, Technische Universität München, University of Bayreuth.